\documentclass[prb,aps,superscriptaddress,twocolumn,amsmath,amssymb,showpacs]{revtex4-2}

\usepackage{amssymb}
\usepackage{booktabs}
\usepackage{multirow}
\usepackage{epstopdf}
\usepackage{float}
\usepackage{graphicx}
\usepackage{dcolumn}
\usepackage{bm}
\usepackage{color}
\usepackage{ulem}
\hfuzz=\maxdimen
\begin{document}
	
\title{Experimental electronic phase diagram in a diamond-lattice antiferromagnetic system}

\author{Liang-Wen Ji} \email[]{lwji@zju.edu.cn}
\affiliation{School of Physics, Zhejiang University, Hangzhou 310058, China}

\author{Wu-Zhang Yang}
\affiliation{School of Sciences, Westlake Institute for Advanced Study, Westlake University, Hangzhou 310064, China}

\author{Yi-Ming Lu}
\affiliation{School of Physics, Zhejiang University, Hangzhou 310058, China}

\author{Jia-Yi Lu}
\affiliation{School of Physics, Zhejiang University, Hangzhou 310058, China}

\author{Jing Li}
\affiliation{School of Physics, Zhejiang University, Hangzhou 310058, China}
	
\author{Yi Liu}
\affiliation{School of Physics, Zhejiang University, Hangzhou 310058, China}
\affiliation{Department of Applied Physics, Zhejiang University of Technology, Hangzhou 310023, China}

\author{Zhi Ren}
\affiliation{School of Sciences, Westlake Institute for Advanced Study, Westlake University, Hangzhou 310064, China}
	
\author{Guang-Han Cao} \email[]{ghcao@zju.edu.cn}
\affiliation{School of Physics, Zhejiang University, Hangzhou 310058, China}
\affiliation{Interdisciplinary Center for Quantum Information, and State Key Laboratory of Silicon and Advanced Semiconductor Materials, Zhejiang University, Hangzhou 310058, China}
\affiliation{Collaborative Innovation Centre of Advanced Microstructures, Nanjing University, Nanjing, 210093, China}
	
\date{\today}

	\begin{abstract}
	We report Ni-doping effect on the magnetic and electronic properties of thiospinel Co$_{1-x}$Ni$_x$[Co$_{0.3}$Ir$_{1.7}$]S$_4$ (0 $\leq x \leq$ 1). The parent compound Co[Co$_{0.3}$Ir$_{1.7}$]S$_4$ exhibits antiferromagnetic order below $T_\mathrm{N} \sim$ 292 K within the $A$-site diamond sublattice, along with a narrow charge-transfer gap. Upon Ni doping, an insulator-to-metal crossover occurs at $x \sim$ 0.35, and the antiferromagnetism is gradually suppressed, with $T_\mathrm{N}$ decreasing to 23 K at $x =$ 0.7. In the metallic state, a spin-glass-like transition emerges at low temperatures. The antiferromagnetic transition is completely suppressed at $x_\mathrm{c} \sim$ 0.95, around which a non-Fermi-liquid behavior emerges, evident from the $T^\alpha$ temperature dependence with $\alpha \approx$ 1.2-1.3 in resistivity and divergent behavior of $C/T$ in specific heat at low temperatures. Meanwhile, the electronic specific heat coefficient $\gamma$ increases substantially, signifying an enhancement of the quasiparticle effective mass. The magnetic phase diagram has been established, in which an antiferromagnetic quantum critical point is avoided at $x_\mathrm{c}$. Conversely, the observed glass-like tail above the critical concentration aligns more closely with theoretical predictions for an extended region of quantum Griffiths phase in the presence of strong disorder.
    \end{abstract}


\maketitle
\section{\label{sec:level1}Introduction}

Thiospinel compounds $AB_2$S$_4$ have been extensively studied over the past few decades~\cite{1-tsurkan-2021}, displaying intricate phenomena such as superconductivity~\cite{2-hagino-1995,3-jin-2021}, non-Fermi-liquid (NFL) behavior~\cite{4-huang-2024}, metal-to-insulator transition~\cite{5-radaelli-2002}, and orbital-glass state~\cite{6-fichtl-2005}. In particular, the complexity of magnetic thiospinels is intensified by frustration effects stemming from lattice geometry and competing exchange interactions~\cite{7-harris-1996,8-balents-2010,9-bergman-2007}. The $A$-site diamond sublattice with only nearest-neighbor interactions is not geometrically frustrated for nearest-neighbor antiferromagnetic (AFM) ordering, in contrast with the case of $B$-site pyrochlore sublattice~\cite{8-balents-2010}. Nevertheless, competing higher-order neighboring interactions give rise to exotic ground states, including spin-orbital liquid near quantum criticality in FeSc$_2$S$_4$~\cite{10-plumb-2016}, spiral-spin liquid and antiferromagnetic skyrmions in MnSc$_2$S$_4$~\cite{11-gao-2017,12-gao-2020}, rendering $A$-site magnetic spinels of tremendous interest.

Recently, a new $A$-site magnetic spinel Co[Co$_{0.3}$Ir$_{1.7}$]S$_4$ was discovered, featuring a high N\'{e}el temperature ($T_\mathrm{N}$) of 292 K~\cite{13-Ji}. Measurements of electrical resistivity and absorption spectrum show a narrow energy gap, which emerges between Ir-$t_\mathrm{2g}$ orbitals and the antibonding states of Co-$t_2$ and S-$p$ orbitals. The narrow charge-transfer gap enables electron hopping between the unoccupied antibonding states of Co-$t_2$ and S-$p$ orbitals via Ir-$t_\mathrm{2g}$ states, thereby facilitating robust nearest-neighbor-coupled antiferromagnetism between the Co($A$) ions with little magnetic frustrations. 

AFM insulators can typically be manipulated by carrier doping, which adjusts the band filling and exerts a profound influence on both transport and magnetic properties~\cite{14-imada-1998,15-khomskii-2014}. For example, Ni doping in iron-pnictide systems introduces excess itinerant electrons and suppresses the AFM order, leading to unconventional superconductivity (SC)~\cite{16-cao-2009,17-li-2009}. Likewise, Ni doping in Co[Co$_{0.3}$Ir$_{1.7}$]S$_4$ can effectively introduce electron carriers. More importantly, electron doping will drive the antibonding states of Co-$t_2$ and S-$p$ orbitals to dominate the Fermi level, which seems to fulfill the electronic ‘gene’ for high-temperature SC~\cite{18-hu-2017,19-hu-2018}. Thus, electron doping via Ni substitution may suppress the antiferromagnetism and trigger exotic phenomena in the cobalt thiospinel system.

\begin{figure*}[t]
	\includegraphics[width=14.5cm]{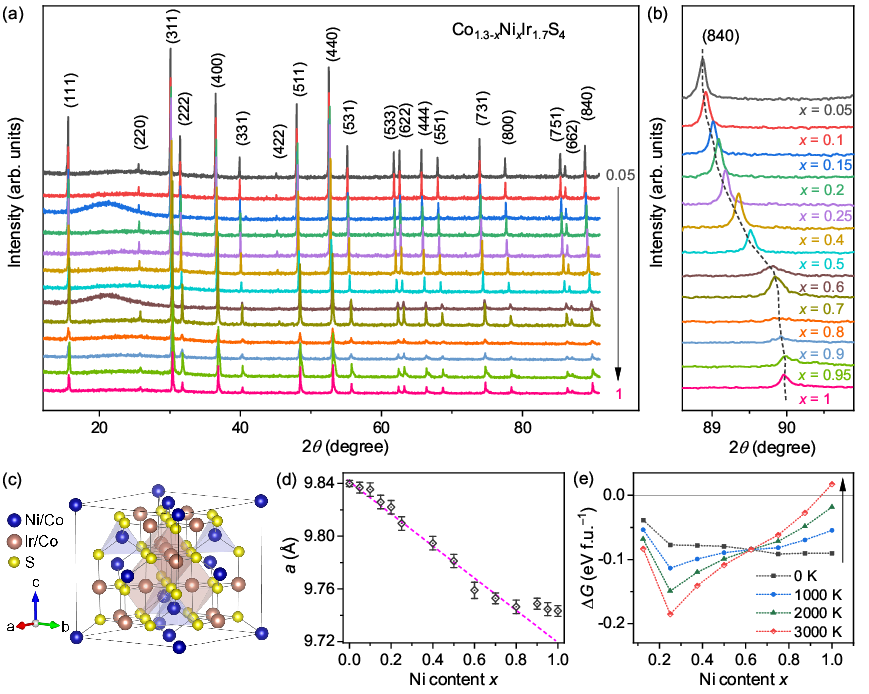}
	\caption{(Color online) (a) Powder x-ray diffraction of Co$_{1.3-x}$Ni$_x$Ir$_{1.7}$S$_4$. (b) Diffraction peak of the (840) crystal plane. (c) Crystal structure of Co$_{1.3-x}$Ni$_x$Ir$_{1.7}$S$_4$. (d) Lattice parameter $a$ as a function of Ni content. The dashed line is a guide to the eye. (e) Gibbs free energy difference between Co$_{8-x}$Ni$_x$[Co$_2$Ir$_{14}$]S$_{32}$ (Co occupies $B$ sites in prior, denoted as $G_1$) and Co$_{8-x+y}$Ni$_{x-y}$[Ni$_y$Co$_{2-y}$Ir$_{14}$]S$_{32}$ (Ni occupies $B$ sites in prior, denoted as $G_2$), $\triangle G(T)$ = $G_1(T) - G_2(T)$ with 1 $\leq x \leq$ 8 and 1 $\leq y \leq$ 2.}
	\label{XRD}
\end{figure*}

In this study, we systematically investigated the effect of Ni substitution for Co($A$) in Co$_{1-x}$Ni$_x$[Co$_{0.3}$Ir$_{1.7}$]S$_4$ (abbreviated as Co$_{1.3-x}$Ni$_x$Ir$_{1.7}$S$_4$) through x-ray diffraction (XRD), electrical transport, magnetic susceptibility and heat capacity measurements. Note that this is the first systematic study of electron doping in the diamond-lattice antiferromagnet Co$_{1.3}$Ir$_{1.7}$S$_4$. With increasing Ni concentration, an insulator-to-metal crossover (IMC) is induced and the AFM transition is gradually suppressed. Accompanied by the disappearance of AFM order, the resistivity and specific-heat data exhibit NFL behaviors at low temperatures, yet without SC. The phase diagram for Co$_{1.3-x}$Ni$_x$Ir$_{1.7}$S$_4$ (0 $\leq x \leq$ 1) has been mapped out, suggesting a possible quantum Griffiths phase (QGP) at $x \textgreater 0.95$. 

\section{\label{sec:level2}Methods}

Polycrystalline samples of Co$_{1.3-x}$Ni$_x$Ir$_{1.7}$S$_4$ (0 $\leq x \leq$ 1) were synthesized via a solid-state reaction using source materials of high-purity Ni powders (99.9 \%), Co powders (99.9 \%), Ir powders (99.95 \%), and S powders (99.99 \%). The nominal stoichiometric mixture of source materials was placed into a sealed evacuated quartz ampoule and slowly heated to 1223-1473 K for two days, depending on the Ni-doping level. The more Ni is introduced, the higher reaction temperature is needed. The resultant powder were reground and pressed, and then were heated to the same temperature for another two days. Note that all the operations of sample handling were carried out in a glove box filled with high-purity argon.

Powder XRD was carried out using a PANalytical x-ray diffractometer with monochromatic Cu-$K_{\alpha1}$ radiation at room temperature. The lattice parameters were obtained by a least-squares fit of reflections in the range of $10^\circ \leq 2\theta \leq 90^\circ$. The dc magnetization was carried out on a Quantum Design Magnetic Property Measurement System (MPMS3). The temperature dependence of electrical resistivity and heat capacity was measured on a Quantum Design Physical Property Measurement System (PPMS-9). A standard four-electrode method was employed for the resistivity measurement. The heat capacity was measured by a thermal relaxation method.

\begin{figure*}[t]
	\centering
	\includegraphics[width=17.5cm]{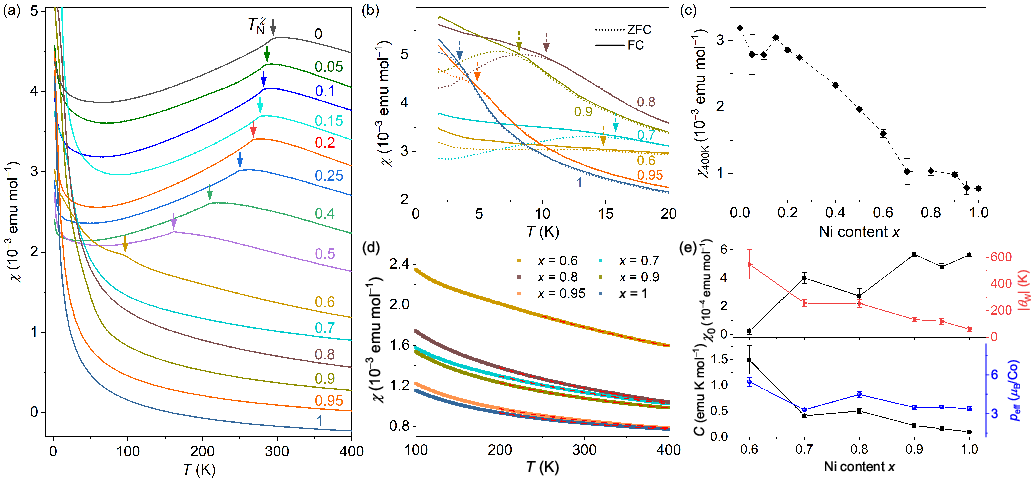}
	\caption{(Color online) (a) Temperature dependence of magnetic susceptibility for Co$_{1.3-x}$Ni$_x$Ir$_{1.7}$S$_4$ with 0 $\leq x \leq$ 1. The arrows represent the antiferromagnetic transition. The data are offset individually for clarity. (b) Temperature dependence of magnetic susceptibility in zero-field-cooling (ZFC) and field-cooling (FC) modes for Co$_{1.3-x}$Ni$_x$Ir$_{1.7}$S$_4$ (0.6 $\leq x \leq$ 1), in which dashed arrows denote the spin-glass-like transition. (c) Magnetic susceptibility at 400 K ($\chi_\mathrm{400 K}$) as a function of Ni content. (d) Magnetic susceptibility of Co$_{1.3-x}$Ni$_x$Ir$_{1.7}$S$_4$ (0.6 $\leq x \leq$ 1) fitted with the Curie-Weiss formula. (e) The upper panel plots the temperature-independent term $\chi_0$ (left axis) and the Weiss temperature $\theta_\mathrm{W}$ (right axis) as functions of Ni content $x$. The lower panel plots the Curie constant $C$ (left axis) and corresponding effective magnetic moments per cobalt (right axis) as functions of Ni content $x$.}
	\label{Mag}
\end{figure*}

The density functional theory (DFT) calculations were performed using the Vienna Ab initio Simulation Package (VASP)~\cite{20-VASP} with the projector augmented-wave method~\cite{21-PAW}. The exchange-correlation energy was calculated with a Perdew-Burke-Ernzerhof (PBE) type functional~\cite{22-GGA}. The wave functions were expanded in the plane waves basis with an energy cutoff of 560 eV. A $8 \times 8 \times 8$ and $6 \times 6 \times 6$ $\Gamma$-centered \textbf{k} mesh was employed for calculations with the 14-atom primitive cell and 56-atom conventional cell, respectively. Conventional cells with varying Ni and Co occupancies were constructed to calculate their total enthalpy. Virtual crystal approximation (VCA) method was adopted to elucidate the change in the electronic structure induced by Ni doping~\cite{23-bellaiche-2000}. The nonmagnetic configurations were adopted in our calculations. All the crystal lattice and ionic sites of the primitive cell and conventional cell were fully relaxed. 

\section{\label{sec:level3}Results and Discussion}

\subsection{\label{sec:level3.1}Crystal structure}

The XRD patterns of the series samples of Co$_{1.3-x}$Ni$_x$Ir$_{1.7}$S$_4$ are shown in Fig.~\ref{XRD}(a). Upon close inspection of the XRD pattern, it is evident that Ni doping leads to obvious shifts in the diffraction peaks (Fig.~\ref{XRD}(b)). Nevertheless, the XRD reflections can be well indexed with the spinel structure in all Ni-concentrations. No secondary phase can be detected. 

The dependence of unit-cell parameter on Ni content is shown in Fig.~\ref{XRD}(d). The lattice constant $a$ decreases from 9.840 \AA\ for $x$ = 0 to 9.746 \AA\ for $x$ = 0.8 at room temperature, showing a shrinkage of the lattice. In both normal NiCo$_2$S$_4$ and inverse Ni$_2$CoS$_4$ thiospinels, Co atoms preferentially occupy the $B$ site ~\cite{24-knop-1968,25-huang-1971}, which is consistent with crystal field stabilization energies~\cite{26-muller-2007}. To further clarify the preferences of Ni and Co for tetrahedral or octahedral sites in this system, we calculated the free energies of several specific configurations with different Ni and Co distributions (Table S1 in the SM~\cite{27-SM}). As shown in Fig.~\ref{XRD}(e), in most cases except for $x \textgreater$ 1, Ni residing at the $A$ site and Co at the $B$ site is indeed more stable. Therefore, substituting smaller Ni$^{2+}$ (0.55 \AA) for Co$^{2+}$ (0.58 \AA) at $A$ site is expected to result in a decrease in unit-cell volume~\cite{28-shannon-1976}. Here, the systematic change in $a$ confirms Ni incorporation into the $A$ site of the lattice. 

In the range of 0.8 $\textless x \leq 1$, however, the change of $a$ deviates from a linear behavior. It is noted that the effective ionic radii for octahedrally coordinated Ni$^{3+}$ (0.56 \AA) is bigger than that of Co$^{3+}$ (0.545 \AA)~\cite{28-shannon-1976}. Meanwhile, in the high-doping region, the free energy difference $\Delta G$ between Co$_{8-x}$Ni$_x$[Co$_2$Ir$_{14}$]S$_{32}$ and Co$_{8-x+y}$Ni$_{x-y}$[Ni$_y$Co$_{2-y}$Ir$_{14}$]S$_{32}$ is minimized due to the increasing entropy contribution, implying a propensity toward site inversion. These phenomena suggest the disordered occupancy of Co and Ni at high doping levels. Furthermore, the deviation of $a$ from the linear fit in Fig.~\ref{XRD}(d) can be utilized to estimate the site inversion degree, which is approximately 4.6 \% to 6.7 \% for $x \geq$ 0.9 (Table S2 in the SM~\cite{27-SM}). 

\begin{figure*}[t]
	\centering
	\includegraphics[width=17.5cm]{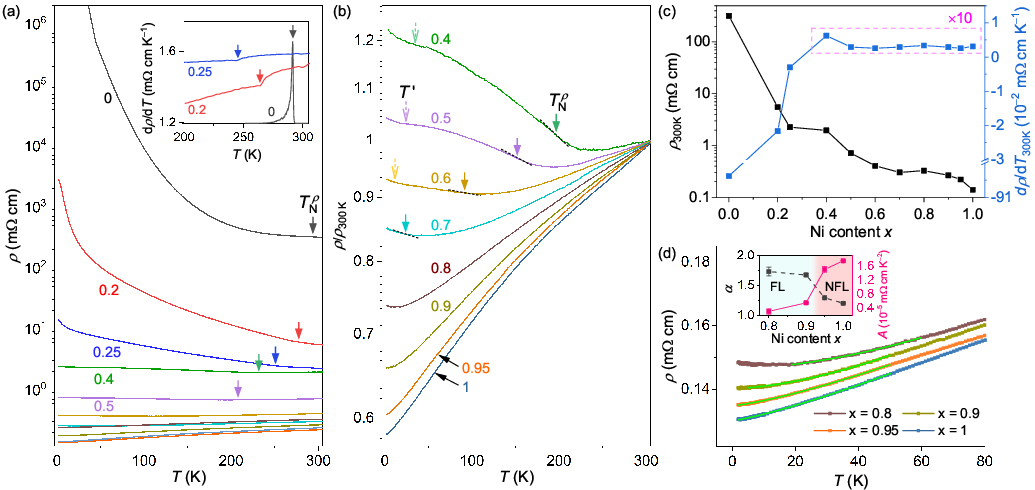}
	\caption{(Color online) (a) Temperature dependence of electrical resistivity for the Co$_{1.3-x}$Ni$_x$Ir$_{1.7}$S$_4$ ploycrystalline samples. The inset is the corresponding d$\rho$/d$T$ in the temperature range 200 to 310 K with an offset. (b) Temperature dependence of normalized resistivity of Co$_{1.3-x}$Ni$_x$Ir$_{1.7}$S$_4$ (0.4 $\leq x \leq$ 1). The arrows and dashed arrows represent the antiferromagnetic ordering and low-temperature anomaly, respectively. (c) Resistivity at room temperature (left axis) and its derivate to temperature at 300 K (right axis) as functions of Ni content. The data in the dashed box are magnified by a factor of 10. (d) The low-temperature resistivity is fitted with the power law $\rho = \rho_0 + A^\prime T^\alpha$, with an offset applied for clarity. The inset plots exponent $\alpha$ (left axis) and coefficient $A$ of the $T$-square term in $\rho = \rho_0 + A T^2$ (right axis) as functions of Ni content. Here, FL and NFL denotes Fermi liquid and non-Fermi liquid, respectively.}
	\label{Res}
\end{figure*}

\subsection{\label{sec:level3.2}Magnetic properties}

Figure~\ref{Mag}(a) shows the magnetic susceptibility $\chi(T)$ of Co$_{1.3-x}$Ni$_x$Ir$_{1.7}$S$_4$ samples. On cooling below room temperature, the parent compound Co$_{1.3}$Ir$_{1.7}$S$_4$ exhibits a peak at $T_\mathrm{N}^\chi \sim$ 292 K. With Ni doping, the $\chi(T)$ data exhibits an overall decline, as evidenced by the decrease of $\chi_\mathrm{400K}$ in Fig.~\ref{Mag}(c). In the low-doping region ($x \leq$ 0.6), this AFM transition peak shifts to lower temperatures, indicating that the N\'{e}el temperature decreases gradually with increasing Ni content. As $x$ increases to 0.6, $T_\mathrm{N}^\chi$ further reduces to $\sim$95 K, which is about 1/3 of the value for pristine Co$_{1.3}$Ir$_{1.7}$S$_4$. While at $x \geq$ 0.7, no signature of AFM transition is observed, and the $\chi(T)$ of Co$_{1.3-x}$Ni$_x$Ir$_{1.7}$S$_4$ exhibits only Curie-Weiss paramagnetic behavior.

Upon careful examination of the low-temperature data, $\chi(T)$ in zero-field-cooling and field-cooling modes bifurcate at $x \geq$ 0.6, as illustrated in Fig.~\ref{Mag}(b). This bifurcation behavior is reminiscent of a spin-glass-like transition, which is commonly observed in $A$-site magnetic spinels~\cite{29-poole-1982,30-naka-2023}. Here, we define the bifurcation point as the spin-glass (SG) transition temperature $T_\mathrm{SG}$, which also decreases with increasing $x$. Given a metallization at $x \geq$ 0.6 (see below), this SG-like behavior at low temperatures can be understood in terms of Ruderman–Kittel–Kasuya–Yosida (RKKY)-induced frustration~\cite{30-naka-2023}.

At 0.6 $\leq x \leq$ 1, the paramagnetic behavior of Co$_{1.3-x}$Ni$_x$Ir$_{1.7}$S$_4$ above 200 K (or 250 K) can be analyzed by the modified Curie-Weiss law, $\chi = \chi_0 + C/(T-\theta_\mathrm{W})$ (Fig.~\ref{Mag}(d)). The fitted parameters as functions of Ni content are shown in Fig.~\ref{Mag}(e). The temperature-independent term $\chi_0$ increases with increasing Ni concentration, consistent with the nearly unchanged core-electron diamagnetism and increasing Pauli paramagnetism ($\chi_\mathrm{P} = \mu^2_\mathrm{B}N(E_\mathrm{F})$, see below). The absolute value of the Weiss temperature $|\theta_\mathrm{W}|$ decreases gradually, indicating the diluted magnetic interaction with Ni doping. The Curie constant $C$ also shows a decreasing trend, which can be basically explained by the nonmagnetic (NM) nature of Ni ions at the $A$ site. Numerically, the decrease of $\chi_\mathrm{400K}$ should be primarily contributed by the reduced Curie-Weiss paramagnetism.

Unlike the NM NiIr$_2$S$_4$~\cite{31-endoh-2001}, Co$_{0.3}$NiIr$_{1.7}$S$_4$ exhibits a notably large magnetic moment of $p_\mathrm{eff}$ $\sim$ 0.8 $\mu_\mathrm{B}$ f.u.$^{-1}$. Considering that only Co$^{2+}$ ions generate magnetic moments, we present the Curie-Weiss paramagnetism signal per Co ion on the basis of the site inversion, as depicted in the lower panel of Fig.~\ref{Mag}(e). At $x =$ 0.6, the local moment per Co ion is $\sim$5.48 $\mu_\mathrm{B}$, comparable to that of pristine Co$_{1.3}$Ir$_{1.7}$S$_4$~\cite{13-Ji}. In the range of 0.6 $\leq x \leq$ 0.9, a general decline in the local moment per Co ion is observed. For $x \geq$ 0.9, the local moment is approximately 3.4 $\mu_\mathrm{B}$/Co, showing little dependence on Ni content. 

\subsection{\label{sec:level3.3}Electrical resistivity}

\begin{figure*}[t]
	\includegraphics[width=14cm]{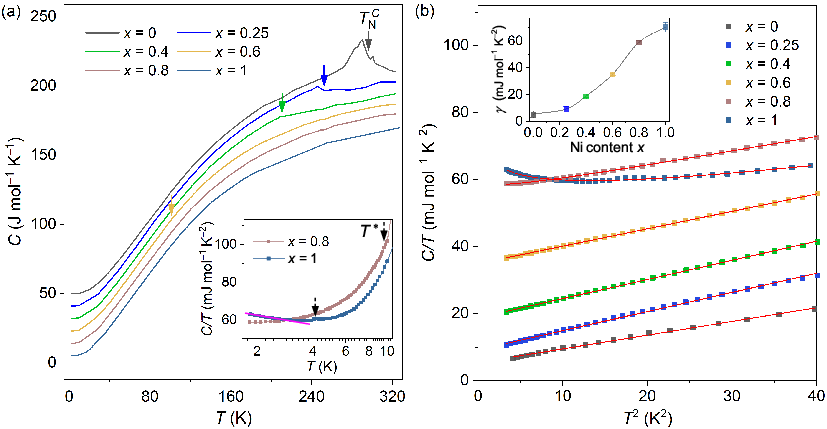}
	\caption{(Color online) (a) Temperature dependence of specific heat $C$ for Co$_{1.3-x}$Ni$_x$Ir$_{1.7}$S$_4$. The data are shifted upward one by one for clarity. The inset shows logarithmic temperature dependence of $C/T$ at low temperatures, with the dashed arrows indicating the anomalies. (b) $C/T$ versus $T^2$ at low temperatures. The inset plots the electronic specific heat coefficient as a function of Ni content.}
	\label{SpecificHeat}
\end{figure*}

Figures~\ref{Res}(a) and \ref{Res}(b) show the temperature dependence of electrical resistivity for Co$_{1.3-x}$Ni$_x$Ir$_{1.7}$S$_4$. The resistivity of Co$_{1.3}$Ir$_{1.7}$S$_4$ exhibits an insulating behavior, with a sharp peak in d$\rho$/d$T$, coincident with the AFM transition at 292 K (inset of Fig.~\ref{Res}(a)). Upon Ni doping, the insulating property is gradually suppressed, as evidenced by $\rho(T)$ plummeting by several orders of magnitude at $x =$ 1. As shown in Fig.~\ref{Res}(c), $\rho_\mathrm{300 K}$ decreases by more than 3 orders of magnitude, falling below 1 m$\Omega$ cm at $x \geq$ 0.5. Concurrently, d$\rho$/d$T$ at room temperature changes sign at $x \approx$ 0.35, suggesting an IMC (Fig.~\ref{Res}(c)). In the insulating state, the resistivity of Co$_{1.3-x}$Ni$_x$Ir$_{1.7}$S$_4$ exhibits a concave anomaly around the AFM transition, allowing one to trace the evolution of magnetic order. 

At $x \geq$ 0.4, the resistivity of Co$_{1.3-x}$Ni$_x$Ir$_{1.7}$S$_4$ exhibits metallic behavior, accompanied by an obvious upturn at low temperatures (Fig.~\ref{Res}(b)). The low-temperature upturn behavior is also observed in other 3$d$ transition-metal compounds, typically attributed to the localization or Kondo effect~\cite{32-rullier-2001,33-zhang-2020}. Notably, the upturn transition of Co$_{1.3-x}$Ni$_x$Ir$_{1.7}$S$_4$ evolves from the Mott-like transition within this system, thus it can be reasonably attributed to the establishment of AFM order. Here, we denote the temperature corresponding to the inflection point of resistivity as $T^{\rho}_\mathrm{N}$, which also decreases as Ni content increases. The AFM ordering persists for Ni concentrations of 0.6 $\leq x \leq$ 0.9. Note that, at 0.4 $\leq x \leq$ 0.6, another upturn emerges at low temperatures, denoted as $T^{\prime}$ and marked by dashed arrows in Fig.~\ref{Res}(b), which coincides with the SG-like transition. The closeness between $T^{\prime}$ and $T_\mathrm{SG}$ suggests that these transitions may originate from similar cobalt spins. That is to say, there should be a re-entrant SG transition below the AFM ordering temperature.

At $x \textgreater$ 0.9, Co$_{1.3-x}$Ni$_x$Ir$_{1.7}$S$_4$ is metallic across the entire temperature range, with no magnetic anomalies observed. On the other hand, the low-temperature resistivity exhibits a power-law behavior (Fig.~\ref{Res}(d)). The power exponent $\alpha$ intriguingly reveals two regimes, as indicated in the inset of Fig.~\ref{Res}(d). At $x \leq$ 0.9, the resistivity exhibits a clear power-law behavior with $\alpha \sim$ 1.7 at low temperatures, which is close to the Fermi-liquid (FL) scenario characterized by $\alpha =$ 2. While at $x \geq$ 0.95, $\alpha$ ranges from 1.2 to 1.3. The significant deviation from $T^2$ dependence suggests NFL behavior in Co$_{1.3-x}$Ni$_x$Ir$_{1.7}$S$_4$, which is rare in spinel~\cite{1-tsurkan-2021,4-huang-2024}. Here we note that, if assuming a Fermi liquid in the low-temperature limit, the fitted coefficient $A$ of the $T^2$ term is found to increase substantially, indicating the increase of effective mass of conduction electrons. 

In chemical substitution systems, disorder plays a crucial role in triggering NFL behavior~\cite{34-miranda-2005}. Particularly in disorder-driven heavy fermion systems, SG freezing is often accompanied by the emergence of NFL behavior~\cite{34-miranda-2005,35-stewart-2001}. In this context, the SG-like behavior observed in magnetic and resistivity measurements suggests a significant impact of disorder. Meanwhile, NFL behavior has not been observed in NiIr$_2$S$_4$ (Figure S1 in the SM~\cite{27-SM}). Thus, the presence of site inversion between Co and Ni atoms may contribute to the exotic NFL in Co-doped NiIr$_2$S$_4$. 

\begin{figure}[t]
	\includegraphics[width=8.5cm]{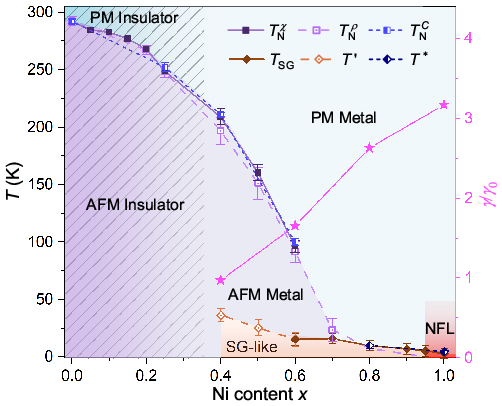}
	\caption{(Color online) Phase diagram of Co$_{1.3-x}$Ni$_x$Ir$_{1.7}$S$_4$ (0 $\leq x \leq$ 1). $T^\chi_\mathrm{N}$ and $T_\mathrm{SG}$ are deduced from magnetic data, $T^{\rho}_\mathrm{N}$ and $T^{\prime}$ are obtained from resistivity data, while $T^{C}_\mathrm{N}$ and $T^{\ast}$ are from heat capacity data. PM Insulator, AFM Insulator, PM Metal, AFM Metal, SG-like, and NFL denote paramagnetic insulator, antiferromagnetic insulator, paramagnetic metal, antiferromagnetic metal, spin-glass-like, and non-Fermi-liquid, respectively. The right axis plots the ratio of the electronic specific-heat coefficient from experimental results ($\gamma$) to DFT calculations ($\gamma_0$).}
	\label{PD}
\end{figure}

\subsection{\label{sec:level3.4}Heat capacity}

Figure~\ref{SpecificHeat}(a) shows the specific heat $C(T)$ for the Co$_{1.3-x}$Ni$_x$Ir$_{1.7}$S$_4$ samples with $x$ = 0, 0.25, 0.4, 0.6, 0.8 and 1. The AFM ordering is confirmed by the peak at $T^{C}_\mathrm{N}$, as remarked by the arrows in Fig.~\ref{SpecificHeat}(a). The N\'{e}el temperature decreases with increasing $x$, consistent with the magnetic and resistive measurements above. Meanwhile, the magnetic related peak progressively weakens with increasing Ni content, suggesting a corresponding reduction in entropy change. This implies a decrease in the local moment, consistent with the fitted Curie constants. As $x$ increases up to 0.8, $C(T)$ shows no signature of AFM transition at high temperatures, but exhibits a low-temperature anomaly (inset of Fig.~\ref{SpecificHeat}(a)), coinciding with the magnetic susceptibility bifurcation shown in Fig.~\ref{Mag}(b). For $x = 1$, the $C/T$ data displays a distinct logarithmic $T$ dependence, as guided by the solid line in the inset of Fig.~\ref{SpecificHeat}(a), aligning with the NFL behavior. The anomaly corresponding to the SG-like transition persists.

Figure~\ref{SpecificHeat}(b) shows the plots of $C/T$ versus $T^2$ at low temperatures. At $0 \leq x \leq 0.6$, a linear relation between $C/T$ and $T^2$ exists, reflecting the dominant contributions from the phonon and AFM magnon part ($\beta T^3$) and the electronic part $\gamma T$. While at higher Ni concentration ($x \textgreater$ 0.8), the $-T\ln T$ contribution from electron-electron interactions should also be taken into consideration~\cite{36-lohneysen-2007}. Therefore, $C/T = \gamma + \beta T^2$ and $C/T = \gamma + \beta T^2 - \eta \ln T$ are used for fitting the low-temperature data of Co$_{1.3-x}$Ni$_x$Ir$_{1.7}$S$_4$ with $0 \leq x \leq 0.6$ and $0.8 \leq x \leq 1$, respectively. 

The dependence of $\gamma$ on Ni content is shown in the inset of Fig.~\ref{SpecificHeat}(b). The electronic specific heat coefficient $\gamma$ increases continuously up to $x = 1$. More importantly, for $x \geq$ 0.6, $\gamma$ of Co$_{1.3-x}$Ni$_x$Ir$_{1.7}$S$_4$ surpasses that of the NM metallic NiIr$_2$S$_4$ (Figure S2 in the SM~\cite{27-SM}). In contrast to NiIr$_2$S$_4$ governed by Fermi-liquid theory, the significantly elevated $\gamma$ of Co$_{1.3-x}$Ni$_x$Ir$_{1.7}$S$_4$ around $x =$ 1 indicates an enhanced effective mass $m^\ast$, as also indicated by the coefficient $A$, which is possibly associated with spin fluctuations near the suppression of magnetic order~\cite{37-michon-2019,38-grossman-2021}. 

\subsection{\label{sec:level3.5}Phase diagram}

With the results above, we mapped out the phase diagram of Co$_{1.3-x}$Ni$_x$Ir$_{1.7}$S$_4$ (Fig.~\ref{PD}), unveiling a fascinating evolution of the electronic and magnetic properties upon Ni doping. Co$_{1.3-x}$Ni$_x$Ir$_{1.7}$S$_4$ alters from an insulator at $x \leq$ 0.25 to a metal at $x \geq$ 0.4, with the IMC occurring at $x \approx$ 0.35. As the Ni concentration increases, the AFM transition is gradually suppressed in both the insulating and metallic states, with the N\'{e}el temperature decreasing from 292 K at $x = 0$ to 23 K at $x = 0.7$. The AFM transition temperatures determined from magnetic susceptibility ($T^\chi_\mathrm{N}$), resistivity ($T^{\rho}_\mathrm{N}$) and specific heat ($T^{C}_\mathrm{N}$) are generally consistent. In the region of $x \geq$ 0.4, a SG-like transition appears at $T^{\prime}$, $T_\mathrm{SG}$ or $T^{\ast}$. All three temperatures ($T^{\prime}$, $T_\mathrm{SG}$ and $T^{\ast}$) decrease with increasing $x$ and coincide with each other. At $x \geq$ 0.95, the AFM ordering vanishes, leaving only the SG anomaly at low temperatures. Notably, NFL behavior emerges within this doping range. 

To further elucidate the anomalous physical properties exhibited, we performed DFT calculations on the electronic structure of Co$_{1.3-x}$Ni$_x$Ir$_{1.7}$S$_4$ with 0.4 $\leq x \leq$ 1 (Figures S3 and S4 in the SM~\cite{27-SM}). Co$_{1.3-x}$Ni$_x$Ir$_{1.7}$S$_4$ remains in a metallic state, with the Fermi energy $E_\mathrm{F}$ locates at a valley in the total density of states (DOS). From $x =$ 0.4 to $x =$ 1, $E_\mathrm{F}$ shift to the right due to the doping of 0.6 extra electrons, resulting in a slight increase of the DOS at $E_\mathrm{F}$. From the bare DOS at $E_\mathrm{F}$ ($N(E_\mathrm{F})$), an electronic specific-heat coefficient of $\gamma_0 = \frac{1}{3}\pi^2k^2_\mathrm{B}N(E_\mathrm{F})$ can be estimated (Fig.~\ref{PD}). At $x =$ 0.4, the derived $\gamma_0$ precisely matches the experimental value. However, the significantly underestimated $\gamma_0$ fails to explain the elevated experimental $\gamma$ for $x \geq$ 0.6. The band structure of Co$_{1.3-x}$Ni$_x$Ir$_{1.7}$S$_4$ exhibits relatively flat dispersions along the X-W line, with narrow bands gradually approaching $E_\mathrm{F}$ as $x$ increases (Figure S4 in the SM~\cite{27-SM}). Thus, the discrepancy between $\gamma_0$ and $\gamma$ may suggest the effective mass renormalization primarily arises from electron-electron interactions~\cite{36-2024-Huang,37-2024-LY}. Particularly, the enhanced electron correlations are crucial for the breakdown of Fermi-liquid in Co$_{1.3-x}$Ni$_x$Ir$_{1.7}$S$_4$~\cite{36-lohneysen-2007}.

In the vicinity of the AFM critical region around $x_\mathrm{c} =$ 0.95, the enhancement of quasiparticle scattering and effective mass implies the possibility of proximity to quantum criticality~\cite{36-lohneysen-2007}. Nonetheless, rather than an AFM quantum critical point (QCP) at the critical concentration, the SG-like transition retains a ‘tail’ for $x \textgreater x_\mathrm{c}$, which is typically observed in disordered systems~\cite{36-lohneysen-2007,39-Thomas-2010}. Notably, the magnetic moment and $\theta_\mathrm{W}$ of Co$_{1.3-x}$Ni$_x$Ir$_{1.7}$S$_4$ remain nonzero up to $x =$ 1, as revealed by magnetic susceptibility. Additionally, the NFL behavior persists in an extended region beyond $x_\mathrm{c}$ (see Figure S1 in the SM~\cite{27-SM}). Resembling the MnSi, Ni$_{1-y}$V$_y$ or Mn$_{1-x}$Fe$_x$Si systems~\cite{40-Pfleiderer-2007,41-Ubaid-2010,42-Mishra-2023}, an infinite-randomness QCP is likely induced by the strong disorder in this system, with a QGP potentially emerging above $x_\mathrm{c}$~\cite{39-Thomas-2010}. The inherent disorder also probably inhibits SC near the quantum critical region~\cite{43-stewart-2017}. 

Within a quantum Griffiths region, power-law singularities of many observables, including magnetic  susceptibility, specific heat, and zero-temperature magnetization are predicted~\cite{44-Castro-1998}. A power-law fit ($T^{-\eta}$ or $H^\eta$) to $\chi(T)$, $M(H)$ at 4 K and $\Delta C/T$ for Co$_{0.3}$NiIr$_{1.7}$S$_4$ yields $\eta$ values of 0.51, 0.8 and 0.42, respectively (Figure S5 in the SM~\cite{27-SM}), which are comparable to those exhibiting NFL behavior due to a QGP~\cite{35-stewart-2001}. In addition, $(M/H)T^\eta$ versus $H/T^\delta$ with $\eta = 0.46$ and $\delta = 1.32$ can generally be collapsed onto a single  curve~\cite{45-Andraka-1991,46-Tsvelik-1993}. Therefore, the subtle interplay of randomness and quantum fluctuations at low temperatures is considered to give rise to the NFL in this system. Further investigations are still needed to ascertain whether quantum Griffiths singularity or related effects are indeed present in this system. 

\section{\label{sec:level4}Concluding Remarks}

We successfully synthesized single-phase samples of thiospinel Co$_{1.3-x}$Ni$_x$Ir$_{1.7}$S$_4$ (0 $\leq x \leq$ 1). The gradual incorporation of Ni at the Co($A$) site leads to an IMC around $x =$ 0.35, the emergence of SG-like behavior at $x \geq$ 0.4 and the suppression of AFM order at the critical concentration of $x_\mathrm{c} \approx$ 0.95. Subsequent to the collapse of AFM order, NFL behavior and a substantial increase of the effective mass are observed. Nevertheless, the phase diagram of Co$_{1.3-x}$Ni$_x$Ir$_{1.7}$S$_4$ reveals an extended SG-like tail beyond $x_\mathrm{c}$, with properties possibly indicative of a QGP at low temperatures due to the randomness generated by strong disorder. The details of quantum Griffiths phase above $x_\mathrm{c}$ will need to be further resolved. Overall, our results provide insights into magnetic quantum criticality with the interplay of quenched disorder in this correlated Co-based diamond-lattice antiferromagnets. 

\begin{acknowledgments}
	
This work was supported by the National Key Research and Development Program of China (2022YFA1403202, 2023YFA1406101), and the National Natural Science Foundation of China (12050003), and the Key Research and Development Program of Zhejiang Province, China (2021C01002). 

\end{acknowledgments}

\section*{References}

\normalem

\bibliographystyle{apsrev4-2}

%

\end{document}